\documentstyle[times,pramana,epsf,graphicx,floats,amssymb]{ias}

\def\plotone#1{\centering \leavevmode
\epsfxsize= 0.6\columnwidth \epsfbox{#1}}

\def\lsim{\mathrel{\rlap{\lower4pt\hbox{\hskip1pt$\sim$}}
    \raise1pt\hbox{$<$}}}                
\def\gsim{\mathrel{\rlap{\lower4pt\hbox{\hskip1pt$\sim$}}
    \raise1pt\hbox{$>$}}}                

\begin{document}
\mark{{ICGC04 Cosmology Workshop }{T. Souradeep }}
\title{Summary of ICGC04 Cosmology Workshop }

\author{ Tarun Souradeep } \address{Inter-University Centre for
Astronomy and Astrophysics,\\ Post Bag 4, Ganeshkhind, Pune 411~007,
India.}

  \pacs{2.0}

\abstract{ Cosmology is passing through a golden phase of rapid
advance. The cosmology workshop at ICGC-2004 attracted a large number
of research contributions from diverse topics of cosmology. I attempt
to classify and summarize the research work and results of the oral
and poster presentations made at the meeting.}

\maketitle


\section{Introduction} 

Recent developments in Cosmology has been largely driven by huge
improvement in quality, quantity and the scope of cosmological
observations. While on one hand, the observations have constrained
theoretical scenarios and models more precisely, some of these
observations have thrown up new challenges to theoretical
understanding and others that have brought issues from the realm of
theoretical speculation to observational verification.  The
contribution to the workshop on cosmology at the ICGC-04 reflects this
vibrancy in the field.

In this article, I summarize both the oral and poster presentations
made at the workshop. The two invited talks at the cosmology workshop
by {\bf S. Majumdar} and {\bf M. Kaplinghat} are included as separate
articles in this issue.  The contributions have been divided into five
broad themes. A brief introduction to each of the themes in the
context of the ICGC04 contributions is given in this section. Sections
2 to 6, summarize the contributions in each of the following five
themes:


{\sl\bf 1. Structure formation:} The formation of large scale
structure in universe has been a very important aspect of modern
cosmology. The `standard' model of cosmology must not only explain the
dynamics of homogeneous background universe, but also (eventually)
satisfactorily describe the perturbed universe -- the generation,
evolution and finally, the formation of large scale structures in the
universe.  Recently, large redshift surveys such as the Las Campanas
Redshift survey (LCRS), $2$degree field (2dF) and SDSS have mapped out
the distribution of matter in the universe~\cite{lah_sut03}. In
conjunction, with other observations these surveys have contributed to
the community wide effort in cosmological parameter
estimation~\cite{sper_wmap03,max_sdss04}.

There are lot of systematic effects that arise in extracting
cosmological information from redshift surveys. Contributions to
ICGC-04 dealt with some such issues.  The distances to distant
galaxies are measured in terms of redshifts.  It is important to
understand in finer detail the mapping from the redshift space to real
space known as the redshift space distortions. The galaxies are just
tracers of the underlying distribution of dark matter in the
universe. Understanding the `bias' in translating from distribution of
galaxies to the dark matter remains an important area of research.
Finally, to compare observed large scale structures to theory, it is
important to devise statistics that can distinguish true morphological
features from chance alignments in terms of robust statistics.

{\sl \bf 2. Cosmic Microwave background anisotropy: } The cosmic
microwave background has played a crucial role in cosmology. In the
past decade, the measured anisotropy in the temperature of the Cosmic
Microwave background has spearheaded the ongoing transition of
cosmology into a precision science~\cite{hu_dod02}. In particular, the
exquisite measurements of the angular power spectrum of CMB
fluctuations have played a crucial role in identifying an emerging
concordance cosmological model. The high quality of data, the good
theoretical understanding of CMB anisotropy and polarization, and the
relatively unambiguous connection between the two has encouraged a
number of researchers to use the CMB data to probe the universe beyond
estimating a set of cosmological parameters. Contributions in ICGC-04
have observationally addressed theoretical assumptions such as the
statistical isotropy of CMB fluctuations, scale invariance in the
primordial power spectrum and parity conservation in electromagnetic
interactions. Efforts to mine the CMB deeper pose data analysis
challenges in particular to account for finer systematic effects in
observations, such as accounting for the non-circularity of the
experimental beam response function presented in the meeting.

{\sl \bf 3. Dark energy:} Observations of the luminosity distance
using high redshift Supernova Ia have indicated that the present
expansion rate of the universe may be accelerating. Within the
Friedman-Robertson-Walker cosmology ( and general relativity), this
implies that the present matter content of the universe is dominated
by dark energy -- a yet unidentified, exotic matter with negative
pressure. Modeled as an ideal hydrodynamic fluid with equation of
state $w$, ($p=w \rho$) the acceleration of universe implies that $w
<-1/3$. Phenomenologically, the simplest model of dark energy is the
cosmological constant (or, vacuum energy) where $w\simeq -1$ is
consistent with data.  Concordant results are also obtained from the
formation of large scale structures in the universe by combining the
exquisite measurements of the angular spectrum of CMB anisotropy with
recent measurements of power spectrum of density perturbation from
large redshift survey. Various combinations of the CMB, High-$z$ SN1a
and galaxy redshift survey data constrain $w\lsim -0.8$.  If the
strong energy condition $w \geq -1$ is not imposed as prior then the
likelihood appears to spill over and peak in the $w<-1$ regime.  Dark
energy with $w <-1$, is rather unusual, if interpreted in terms of a
hydrodynamic fluid or scalar field energy density. A number of
contributions in the workshop are theoretical models and scenarios
attempting to explain dark energy with $w < -1$.  An interesting
possibility which is explored in a contribution to this meeting is is
to recover the evolution of the equation of state, $w$, with redshift
from the data without imposing the strong energy condition prior.

{\sl\bf 4. Early Universe \& extra dimensions:} In the hot Big Bang
scenario, the present universe is inescapably linked to the ultra high
energy physics of the early universe. It is fair to say that cosmology
will remain incomplete without adequate understanding of the early
universe. The initial singularity of big bang remains an enigma that
time and again attracts imaginative solutions. The recent proposal to
'cap' the initial FLRW model with a static Einstein universe is
explored in the context of higher derivative extensions of gravitation
in one of the workshop contributions.  The primordial perturbations
believed to be generated during inflation is one of most promising
probe of physics at ultra-high energies, possibly, even up to
trans-Planckian energy scales. A contribution here has explored the
signature of trans-Planckian physics that respects Lorentz symmetry.
Besides scalar density perturbations, the gravity wave background from
inflation is an important clue to the early universe physics.  For
example, overproduction of gravity waves can severely constrain some
braneworld inflation models with steep inflaton potential, unless, as
shown one of the contributions, the mechanism of reheating is
reworked.  Although cosmic defects produced during phase transition in
the early universe seem unlikely to be the dominant source of
primordial perturbations, the possibility that they play a
sub-dominant role in structure formation is still an open
possibility. A formalism for studying perturbations from defects was
presented in the workshop.

In the past few years, the possibility of large extra dimensions has
caught the fancy of both theoretical high energy physicists as well as
cosmologists. These braneworld scenarios, where the observed $3+1$
dimensions and all interactions other than gravity reside on a brane
embedded in a higher dimension, are usually motivated by string theory.
These scenarios, initially invoked to address the hierarchy problem of
disparity between the SUSY breaking and the Planck scale, have been
used and studied in a variety of other contexts. The observed
accelerating universe and dark energy have been linked to braneworld
scenario. Construction of stable brane configurations that may do this
was discussed in the meeting. Braneworld scenarios can have
interesting consequences for inflation as shown by some contributions
in the meeting.

{\sl\bf 5. Alternative approaches: } Despite the emergence of
concordant cosmological model, there is enough width left in cosmology
for exploring radically different ideas. The reason is that recent
advances in cosmology have been more on the phenomenological, rather
than conceptual aspects.  Alternative ideas may be the key to some of
the intriguing puzzles of contemporary cosmology. While many of these
ideas face increasingly difficult challenges to match observations, it
is important to allow them to be judged by observations, and not
theoretical prejudices.


\section{ Large scale structures in the universe}

\begin{figure}[t]
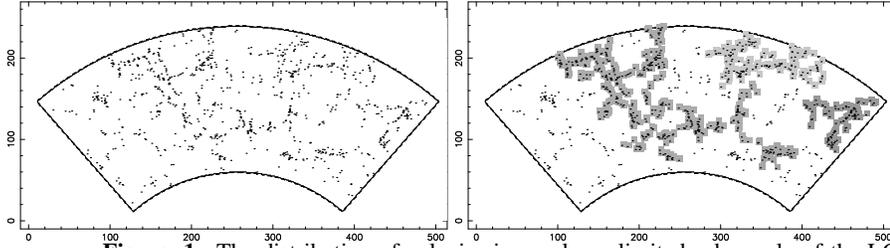

\begin{center}
\includegraphics[scale=0.25,angle=-90]{pg.0.ps}
\includegraphics[scale=0.25,angle=-90]{pg.4.ps}
\caption{The distribution of galaxies in a volume limited subsample of
the LCRS slice at declination -3 is shown in the figure on
the left. The figure on the right shows a coarse grained distribution
of galaxies.  Some of the largest clusters are identified and
highlighted. These filamentary clusters are can be shown to be
statistically significant.}
\end{center}
\label{fig:som}
\end{figure}

The morphology of the structures in the large scale distribution of
matter in the universe seen in the recent redshift surveys has to be
quantified with a statistically robust measures. One of the most
striking visual features in redshift surveys is that galaxies appear
to be distributed along filaments that are interconnected to form a
network which extends across the entire survey like a cosmic
web. Figure~\ref{fig:som} shows the distribution of galaxies in one of
the slices of the Las Campanas redshift Survey (LCRS) and highlights
the filamentary structures that appear within it.  {\bf S. Bharadwaj}
reported a novel analysis (with S. Bhavsar and J. Sheth) of the Las
Campanas redshift Survey (LCRS) to determine the extent to which the
filaments are genuine, statistically significant features as against
the possibility of their arising from chance
alignments~\cite{bhar04}. They find that one of the LCRS slices has
statistically significant filamentary features spanning scales as
large as $70$ to $80\,h^{-1}{\rm Mpc}$, whereas filaments spanning
scales larger than this are not statistically significant (see
fig.~\ref{fig:som}). For the five other LCRS slices, filaments of
lengths $50$ to $70\,h^{-1} {\rm Mpc}$ are statistically significant,
but not beyond. {\em The reality of the $80\,h^{-1}{\rm Mpc}$
filamentary features in the LCRS slice make them the longest coherent
features presently known.}


The galaxy two point correlation function measured from redshift
surveys exhibits deviations from the predictions of linear theory on
scales as large $20\,h^{-1} {\rm Mpc}$ where we expect linear theory
to hold. Any attempt at analyzing the anisotropies in the redshift
correlation function and determining the linear redshift distortion
parameter requires these effects to be taken into account. Usually
these non-linear effects are attributed to galaxy random motions, and
a heuristic model where the linear redshift correlation is convolved
with the random pairwise velocity distribution function is used.  {\bf
S. Bharadwaj} reported investigations (with B. Pandey) of a different
model which is derived under the assumption that linear theory holds
in real space, and which takes into account all non-linear effects
that are introduced by the mapping from real to redshift
space~\cite{bhar_pan04}.  They test this model using N-body
simulations and find that the pairwise velocity dispersion predicted
by all the models considered are in excess of the values determined
directly from the N-body simulations. {\em This indicates a shortfall
in the understanding of the statistical properties of peculiar
velocities and their relation to redshift distortion.}


The distribution of galaxies is believed to trace the underlying
distribution of dark matter quantified by a bias factor.  {\bf
J. Bagla} presented results from a study (with S. Ray) of the moments
of counts in cells for a set of models in real space and redshift
space.  A comparison of the moments for the entire distribution as
well as for regions with over-density above a certain threshold offers
insight into the differences between clustering of galaxies and dark
matter.  They focus mainly on non-linear scales.  Tree-PM simulations
were used to generate the distribution of particles.  They find that
at non-linear scales the bias for the second moment of the
distribution is scale dependent.  The scale dependence is such that
the variation of $\sigma$ with scale is the same for all the models
studied here.  The amplitude of $\sigma$ is very different for these
models, and models with a more negative index have a larger linear
bias.  Skewness for different models studied here is very different
for the entire distribution and has expected values for the power law
models in the linear as well as the extreme non-linear regime.  {\em
Skewness for over-dense regions in all of these models is the same in
redshift space in the non-linear regime}. This startling result
implies that in the non-linear regime the skewness of the distribution
of particles in over-dense regions is not sensitive to initial
conditions.



{\bf S. Ray} reported on the development of a parallel version of the
Tree-PM code optimized for cluster computing.  The Tree-PM method is a
hybrid method for cosmological N-Body simulations that uses both the
tree and mesh based methods to compute force.  The force is divided
into two parts and mesh method is used to compute the long range force
and the tree method is used for computing the short range
force~\cite{bag_ray03}.  They use domain decomposition for
distributing the computation of short range force on a set of
processors.  Functional decomposition is used to assign the tasks for
computing long range and short range forces.  They also introduce
optimizations to reduce the communication overheads. {\em The present
version of the code scales almost linearly up to $34$ processors for
simulations with $1.6 \times 10^7$ particles.  Time taken per step per
particle for these simulations is about $18\mu$s.}  



{\bf J. Prasad} reported initial results on a study (with S. Ray and
J. Bagla) of the interaction of fluctuations at very different scales
by simulating collapse of a plane wave with varying amount of
substructure.  The substructure is modeled as power in a narrow range
of scales at $k_s \gg k_l$, where $k_l$ is the wave number for the
long range wave mode.  The substructure is along three directions
whereas the plane wave collapses along one direction.  In absence of
substructure they find the usual structure where a pancake forms and
has an increasing number of streams as one nears the center.  In
presence of small scale fluctuations, they find that the size of the
multi-stream region shrinks by a small amount as the amplitude of
substructure is increased. They believe that this is due to transfer
of kinetic energy in directions transverse to the plane wave due to
interaction of clumps.  The plane wave influences evolution of small
scale fluctuations strongly.  {\em Formation of clumps is suppressed
in the regions made under-dense by the plane wave, whereas merging in
the over-dense regions leads to a rapid growth of fluctuations at
small scale.}

A study of density perturbations of a cosmological scalar field
addressing the possibility of using the instability mechanism of Jeans
theory, to form self gravitating configurations from a real scalar
field scalar field approach to Jeans mass calculation was presented by
{\bf M. Joy} and V. C. Kuriakose~\cite{mjoy}. They consider a massive
scalar field arbitrarily coupled to a gravitational background, with
the stress-energy tensor expectation values of the quantum field
fluctuations computed in a coherent state.  {\em It is shown that the
self-interaction of the scalar field influences the character of
instability and the value of the Jeans wave number is altered by the
effects of self-interaction.}


The observations of clustering in the distribution of HI can be used
to study large scale structures at high redshift.  {\bf S. Bharadwaj}
presented a study of the possibility of using Giant Meter-wave Radio
Telescope (GMRT) to probe large scale structures in the universe at
high redshift by studying fluctuations in the redshifted 1420 MHz
emission from the neutral hydrogen (HI)~\cite{bhar_srik04}. The study
focuses on the cross-correlations between the visibility signal
measured at different baselines and frequencies in radio
interferometric observations. They show that the visibility
correlations directly probe the power spectrum of HI fluctuation, and
present analytic estimates of the signal expected in two of the GMRT
bands centered at 325 and 610 MHz. They also simulate GMRT
observations including the expected HI signal, galactic and
extragalactic foregrounds and system noise. {\em The preliminary
results indicate that it may be possible to detect the HI signal in
around 1000 hours of observations.}

\section{ Cosmic Microwave Background anisotropy}

\begin{figure}\label{fig:haj}
\plotone{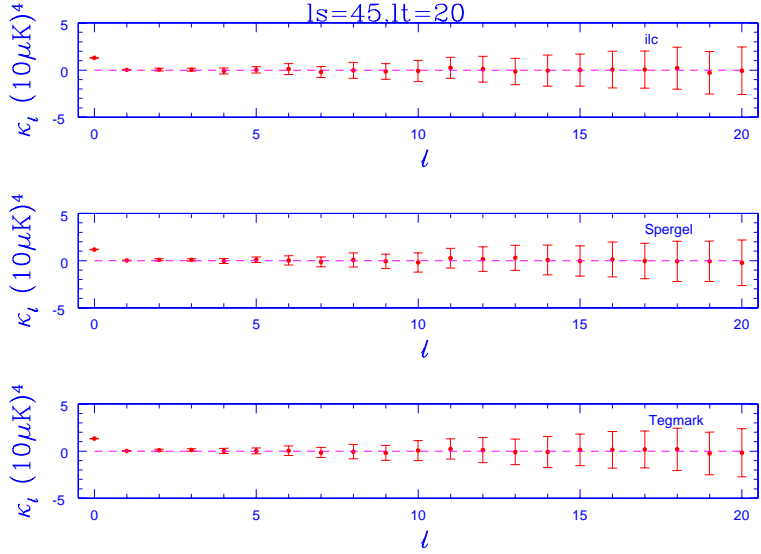}
\caption{The Bipolar power spectrum of three different CMB maps
obtained from the WMAP data filtered by window function to retain
power only on multipole values in the range $20\lsim l\lsim 45$. The
maps appear to be statistically isotropic.}
\end{figure}

The statistical expectation values of the temperature fluctuations of
cosmic microwave background (CMB) are assumed to be preserved under
rotations of the sky. The assumption of statistical isotropy (SI) of
the CMB anisotropy should be observationally verified since detection
of violation of SI could have profound implications for cosmology.
The Bipolar power spectrum (BiPS) has been recently proposed as a
measure of violation of statistical isotropy in the CMB anisotropy
map\cite{haj_sour03}. {\bf A. Hajian} reported results from a BiPS
analysis (with T. Souradeep) statistical isotropy of the CMB
anisotropy maps obtained from the first year of data from the WMAP
satellite. The CMB maps were smoothed by a family of window functions
to isolate and test the SI in the different regions of the multipole
space. Figure~\ref{fig:haj} also shows the BiPS of three different CMB
maps obtained from the WMAP data filtered to retain power with
$20\lsim \ell \lsim 45$.  {\em Preliminary results indicate that the
CMB anisotropy maps from WMAP do not strongly violate statistical
isotropy.}

 
 The non-circularity of the experimental beam has become progressively
important as CMB experiments strive to attain higher angular
resolution and sensitivity.  Recent CMB experiments such as ARCHEOPS,
MAXIMA, WMAP have significantly non-circular beams.  Future
experiments like Planck are expected to be even more seriously
affected by non-circular beams. {\bf S. Mitra} reported a study of the
effect of a non-circular beam on CMB power spectrum estimation (done
with A. Sengupta and T. Souradeep)~\cite{mit04}.  They compute the
bias introduced estimated power spectrum. They construct an unbiased
estimator using the bias matrix.  The covariance matrix of the
unbiased estimator is computed for non-rotating smooth beams. The WMAP
beams maps are fitted and shown to significantly non-circular.  The
effect of a non-circular beam on power spectrum estimate is calculated
for a CMB map made by an experiment with a beam which is non-circular
at a level comparable to the WMAP beam.

\begin{figure}[h]\label{pkrec}
\plotone{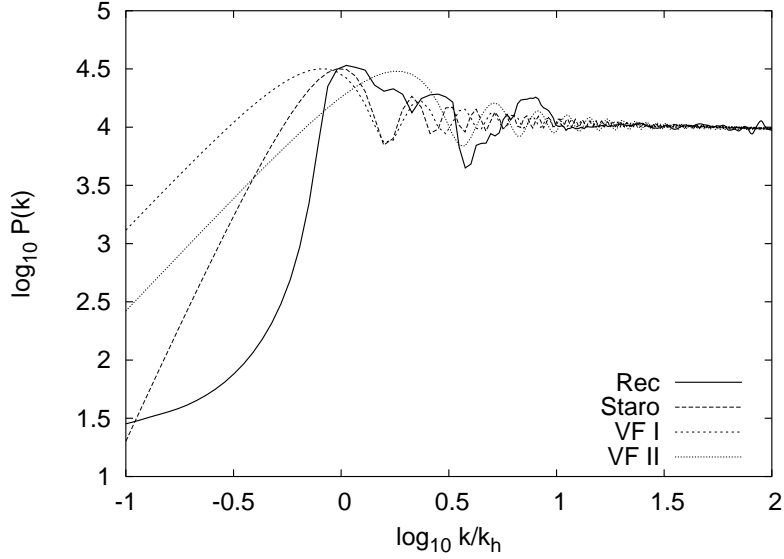}
\caption{The $P(k)$ recovered from the WMAP is plotted (solid).  The
predictions of two simple theoretical scenarios mentioned in the text
that remarkably match the gross features of the infra-red cutoff in
the recovered spectrum are also shown in the figure. }
\end{figure}

Cosmological parameters estimated from CMB anisotropy assume a
specific form for the spectrum of primordial perturbations believed to
have seeded the large scale structure in the universe.  Accurate
measurements of the angular power spectrum $C_l$ over a wide range of
multipoles from the Wilkinson Microwave Anisotropy Probe
(WMAP)~\cite{ben_wmap03} has opened up the possibility to deconvolve
the primordial power spectrum for a given set of cosmological
parameters.  {\bf A. Shafieloo} presented results from a work (with
T. Souradeep) on the direct estimation of the primordial power
spectrum from WMAP measured angular power spectrum of CMB anisotropy
using an improved Richardson-Lucy deconvolution
algorithm~\cite{arm_sour04}.  The most prominent feature of the
recovered $P(k)$, shown as the solid curve in figure~\ref{pkrec}, is a
sharp, infra-red cut off on the horizon scale. It also has a localized
excess just above the cut-off which leads to great improvement of
likelihood over the simple monotonic forms of model infra-red cut-off
spectra considered in the post WMAP literature.  The form of infra-red
cut-off is robust to small changes in cosmological parameters.
Remarkably similar forms of infra-red cutoff is known to arise in very
reasonable extensions and refinements of the predictions from simple
inflationary scenarios. In figure~\ref{pkrec}, the curve labeled
`staro' is the primordial spectrum when the inflaton potential has a
kink-- a sharp, but rounded, change in slope~\cite{star92} and two
curves labeled `VF' are the modification to the power spectrum from a
pre-inflationary radiation dominated epoch~\cite{vilfor82}.

In addition to temperature fluctuations, the CMB photons coming from
different directions have a random, linear polarization. The
polarization of CMB can be decomposed into $E$ part with even parity
and $B$ part with odd parity.  Besides the angular spectrum
$C_l^{TT}$, the CMB polarization provides three additional spectra,
$C_l^{TE}$, $C_l^{EE}$ and $C_l^{BB}$ which are invariant under parity
transformations. The level of polarization of the CMB being about a
tenth of the temperature fluctuation, it is only very recently that
the angular power spectrum of CMB polarization field has been
detected. The Degree Angular Scale Interferometer (DASI) has measured
the CMB polarization spectrum over limited band of angular scales in
late 2002~\cite{kov_dasi02}. The WMAP mission has also detected CMB
polarization~\cite{kog_wmap03}. WMAP is expected to release the CMB
polarization maps very soon.


Parity violating interactions open up the possibility for measuring
non-zero $C_l^{TB}$ and $C_l^{EB}$ power spectra. One such possibility
is a parity-violating interaction of the antisymmetric tensor
Kalb-Ramond (KR) gauge field and the electromagnetic
field~\cite{maj03}. {\bf P. Majumdar} presented results that show that
parity forbidden $C_l^{TB}$ spectra can arise due to such
interactions~\cite{maj04}. The coupling also leads to an effective
time-dependent fine-structure constant in the current cosmological
epoch, pointing thereby to possible correlations between these two
disparate phenomena.



\section{Accelerating universe and Dark energy}

\begin{figure}
\plotone{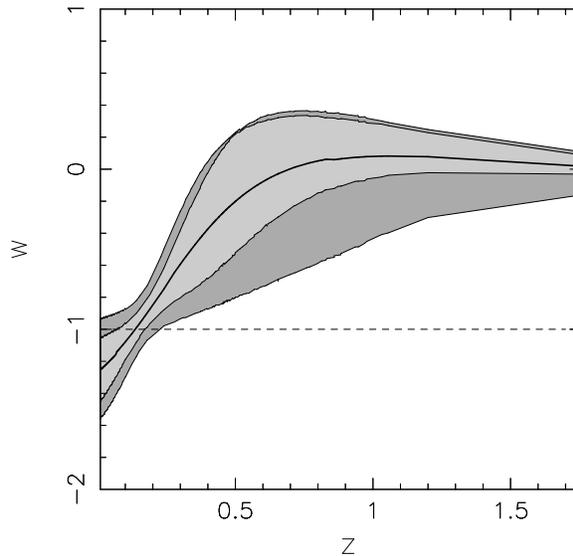}
\caption{
The evolution of $w(z)$ with redshift for $\Omega_{0 \rm m}=0.3$.  The
reconstruction is done using a polynomial fit to dark energy, for the
latest HZT sample of 194 type Ia SNe. The thick solid line shows the
best-fit, the light gray contour represents the $1\sigma$ confidence
level, and the dark gray contour represents the $2\sigma$ confidence
level around the best-fit. The dashed line represents $\Lambda$CDM. No
priors are assumed on $w(z)$.  
}
\label{fig:var_w}
\end{figure}

Improvements in the measurements of luminosity distance as a function
of redshift from High redshift SN1a may eventually allow direct
recovery of the evolution of the equation of state of the dark energy
component in the universe at low redshifts.  {\bf U. Alam} reported
results from a new study (with V. Sahni, T.  D. Saini,
A. A. Starobinsky) where dark energy parameters are reconstructed from
the latest data set of 194 supernovae~\cite{ton03,bar03} without any
priors on the equation of state $w$~\cite{alam}. They find that dark
energy evolves rapidly and metamorphoses from dust-like behavior at
high redshift ($w \simeq 0$ at $z \sim 1$) to a strongly negative
equation of state at present ($w \lsim -1$ at $z \simeq 0$), as shown
in the figure~\ref{fig:var_w}. Dark energy metamorphosis appears to be
a robust phenomenon which manifests for a large variety of supernova
data samples provided one does not invoke the weak energy prior $\rho
+ p \geq 0$. {\em These results indicate that dark energy with a
rapidly evolving equation of state may provide a compelling
alternative to a cosmological constant if data are analyzed in a
prior-free manner.}


Typical scalar field models do not provide a scenario with $w<-1$.  A
radical possibility is to consider a phantom scalar field which has
negative kinetic energy and violates null dominant energy condition --
now popularly referred to as a `phantom field'. {\bf P. Singh},
reported work (with M. Sami, and N. Dadhich) on a model of phantom
field motivated S-brane dynamics using the Supernova Ia observations
to constrain the parameters of the model~\cite{param}.  {\em They find
that the model fits High redshift Supernova data fairly well for a
large range of parameters and favors a $w < -1$.}

The negative kinetic energy term in the Lagrangian of a phantom field
is not very well motivated and also suffers from severe
instability. {\bf S. Das} reported work on explaining dark energy with
$w<-1$ in the Brans-Dicke theory of gravity. {\em The results indicate
that $w<-1$ could be obtained if the gravitational constant is slowly
varying with time in a canonical Brans-Dicke theory of gravity without
conflict with the solar system constraints.}

The de Broglie-Bohm approach to quantum mechanics helps to demonstrate
that the Wheeler-De Witt equation is equivalent to the corresponding
classical equation for a special potential; i.e., the universe is both
quantum and classical at the same time. {\bf M. John} presented a
comparison of prediction of this approach with observations of
High-redshift Supernova data. {\em They claim that the observational
level, this special solution to cosmology is as good as the
conventional matter-$\Lambda$ cosmologies.}

\section{Early universe and Brane world }

Curing the initial singularity of standard FLRW models has been a long
standing endeavor of cosmologists. Recently, Ellis and Maartens
studied a class of cosmological models in which there is no
singularity, no beginning of time and no horizon
problem~\cite{ell_maar04}. The universe starts out as an almost static
universe and expands slowly, eventually evolving into a hot big-bang
era. An example of this scenario is a closed model with a minimally
coupled scalar field, with a special self-interaction potential. This
potential may be obtained after a conformal transformation of the
metric of a higher derivative theory.  {\bf S. Mukherjee} reported a
detailed study (with B. Paul, S. Maharaj and A. Beesham) of higher
derivative theories, including a cosmological constant and a quadratic
term ($R^2$) in the Lagrangian density, where $R$ is the scalar
curvature. The field equations are analyzed to determine the general
characteristics of the evolution and some quantities of cosmological
interest are calculated. The results are compared with those of the
proposed emergent universe. {\em The stability of the results have
been studied by considering an $R^3$ term as a perturbation of the
quadratic Lagrangian density. The possibility of a quantum creation of
the emergent universe in quantum cosmology has also been considered.}


{\bf B. Modak} presented application of Noether symmetry as powerful
tool to find the solution of the field equations for scalar tensor
theory including curvature quadratic term. {\em A few physically
reasonable solutions like power law inflation were presented.}

Quintessential inflation describes a scenario in which both inflation
and dark energy are described by the same scalar field. In
conventional brane world models of quintessential inflation
gravitational particle production is used to reheat the universe. This
reheating mechanism is very inefficient and results in an excessive
production of gravity waves which can even violate nucleosynthesis
constraints and invalidate the model~\cite{sami}. {\bf M. Sami}
described a new method of realizing quintessential inflation on the
brane in which inflation is followed by `instant preheating'. {\em The
larger reheating temperature in this model is shown to result in a
smaller amplitude of relic gravity waves which is consistent with
nucleosynthesis bounds.} The relic gravity wave background has a
`blue' spectrum at high frequencies and is a generic byproduct of
successful quintessential inflation on the brane.

{\bf S. Biswas} presented a study of fermion particle production in
early universe using the complex trajectory WKB method developed
earlier. They study the particle production in periodic potential,
generally used in inflationary cosmology. {\em Using this method, they
recover results obtained in literature earlier, such as
~\cite{kof_green00}.}

Due to the tremendous red-shift that occurs during the inflationary
epoch in the early universe, it has been realized that trans-Planckian
physics may manifest itself at energies much lower than the Planck
energy. The presence of a fundamental scale suggests that local
Lorentz invariance may be violated at sufficiently high energies.
However, certain astrophysical observations seem to indicate that
local Lorentz invariance may be preserved to extremely high energies.
This suggests considering models of trans-Planckian effects that
preserve local Lorentz invariance. {\bf L. Sriramkumar} (with
S.~Shankaranarayanan) presented one such model and evaluated the
spectrum of density perturbations during inflation in the
model~\cite{srir_sank04}. {\em They find that, in the case of
exponential, as well as, power-law inflation, the corrections to the
standard scale-invariant perturbation spectrum (in the Bunch-Davies
vacuum) turn out to be small.}

Several promising models (e.g. D-brane inflation) of high energy
physics inspired inflationary scenarios terminate with the production
of topological defects.  In order to investigate the observational
implications of such models, {\bf G. Amery} described a complete 4-d
synchronous gauge formalism that may include non-zero curvature and
cosmological constants, multiple scalar fields, and topological
defects or other active sources~\cite{amery}.  {\em The formalism
provides a concise and geometrically sound description of
energy-momentum conservation on all scales, from which appropriate
initial conditions may be obtained}.  They present preliminary
investigations of the possible contributions by active sources to the
CMB data.


{\bf B. C. Paul} presented a study of chaotic inflationary universe in
a Brane world model~\cite{paul}.  They study the evolution of the
universe with a minimally coupled self interacting scalar field when
the kinetic energy dominates the potential energy and vice versa and
obtain cosmological solutions which permits inflation. In four
dimensional gravitation (GTR), the initial value of the inflaton field
$\phi_{i} \gsim {\rm few}\, M_{4}$ required for a sufficient
inflation is physically unrealistic. {\em In contrast, they show that
in brane world model sufficient inflation may be obtained even with an
initial scalar field having value less than the usual four dimensional
Planck scale.}


{\bf H.~K.~Jassal} reported on some cosmological consequences of the
five dimensional, two brane Randall-Sundrum brane world scenario.  The
radius of the compact extra dimension is taken to be time
dependent. Integrating over extra dimensions, the four dimensional
action reduces to that of scalar tensor gravity.  The radius of the
extra dimension stabilizes to a nonzero separation of branes very
quickly.  A simple quadratic potential with minimum at zero leads to
stabilization at comparable level but also allows for accelerated
expansion.  After stabilizations the potential does not play any other
role except contributing the dark energy component at late times. {\em
It is shown that requirements for solving the hierarchy problem and
getting an effective dark energy can be satisfied simultaneously.}


\section{Alternate views and ideas in Cosmology}

If confirmed, the often discussed periodicity in the redshift
distribution of quasars may not be readily explained in the standard
Big Bang model.  {\bf P. K. Das} presented results that indicate that
the Variable Mass Hypothesis scenario of Hoyle- Narlikar Theory
redshift quantization can be invoked to explain any observed
periodicity or quantization of quasar redshift distribution.


By considering thermodynamics of open systems in cosmology, Prigogine
has proposed an irreversible matter creation process accompanied with
large- scale entropy production.  {\bf P.~Gopakumar} and
G.~V.~Vijayagovindan discussed the application of this scenario in 3-
brane world cosmology in 5-dimensions. The matter creation rate is
found to affect the evolution of the scale factor both at high and low
energy densities. In the standard brane world scenario, cooling is
much slower than in the FLRW case, at high energy densities. With the
matter creation in the brane world, the standard FLRW evolution is
regained. As a consequence the temperature at the freezing-out of
neutrons to protons ratio is the same as in the standard scenario.


{\bf M. Govender} presented a simple method which generalizes the
static isothermal universe first studied by Saslaw {\em et
al}~\cite{sas}.  The cosmic fluid in the static model obeys a
barotropic equation of state of the form $p= \alpha
\rho$~\cite{waghetal}.  It has been argued that the isothermal
cosmological model of Saslaw {\em et al} could represent the
asymptotic state of the Einstein-de Sitter model.  The generalized
model could describe an isothermal sphere of galaxies in
quasi-hydrostatic equilibrium with heat dissipation driving the system
to equilibrium.  A thermodynamical treatment within the framework of
extended irreversible thermodynamics of the model is carried
out~\cite{maar97}.

\section*{Acknowledgment} 

I thank the Scientific Organizing Committee of ICGC-04 for the
opportunity to chair the ICGC-04 cosmology workshop.  I would like to
thank the local organizing committee for ensuring the smooth running
of the workshop activities and a wonderful time at Kochi.

\end{document}